\documentstyle[aps,pra,epsf]{revtex}
\begin{document}
\title{Insulator-Superfluid transition of spin-1 bosons in an optical lattice in
magnetic field}
\date{\today }
\author{A. A. Svidzinsky and S. T. Chui}
\address{Bartol Research Institute, University of Delaware, Newark, DE 19716}
\maketitle

\begin{abstract}
We study the insulator-superfluid transition of spin-1 bosons in an optical
lattice in a uniform magnetic field. Based on a mean-field approximation we
obtained a zero-temperature phase diagram. We found that depending on the
particle number the transition for bosons with antiferromagnetic interaction
may occur into different superfluid phases with spins aligned along or
opposite to the field direction. This is qualitatively different from the
field-free transition for which the mean-field theory predicts a unique
(polar) superfluid state for any particle number.
\end{abstract}

\pacs{PACS numbers:  03.75.Kk, 03.75.Lm, 03.75.Mn, 32.80.Pj}

\section{Introduction}

The insulator-superfluid transition is an example of quantum phase
transitions which take place at zero temperature. At the transition point
the quantum ground state of the system changes in some fundamental way. This
is accomplished by changing some parameter in the Hamiltonian of the system.
Possible examples are transition between different quantized plateaus in a
quantum-Hall effect under the change of magnetic field, transition between
different phases in high-$T_c$ superconductor compounds under the variation
of doping, metal-insulator transition in a conductor under the change of
disorder, or phase transitions in baryonic matter under the change of
density (e.g., transition from nuclear matter to a uniform liquid of
neutrons, protons and electrons at the interface between a crust and a core
of a neutron star).

Originally the insulator-conductor transition has been studied in Fermi
systems. However, analogous phenomena occur for a system of bosons, for
example, in $^4$He absorbed in porous media or granular superconductors in
which the Cooper pairs may be considered as bosons. For pure Bose systems at
zero temperature the conducting phase is the superfluid, so that the
insulator-conductor transition corresponds to the onset of superfluidity. In
the context of insulator-superfluid transition in liquid $^4$He the problem
has been studied in \cite{Chan88,Fish89}. Recently such a problem attracted
much attention due to experimental realization of Bose-Einstein condensation
(BEC)\ of atoms in magnetic and optical traps. It has been proposed that the
insulator-superfluid transition might be observable when an ultracold gas of
atoms with repulsive interactions is trapped in a periodic optical lattice
\cite{Jaks98}. Recently the transition in $^{87}$Rb condensate in a
three-dimensional simple-cubic potential was realized experimentally by
changing the potential depth \cite{Grei02}. An artificial optical lattice
has a great advantage in that the system is basically defect-free which
provides ideal conditions for study the quantum phase transitions.

In the insulator phase the atoms are localized (by the effect of the
potential) in the lattice sites so that fluctuations in the atom number on
each site are suppressed. In such a state there is no phase coherence across
the lattice. In addition the Mott insulator phase is characterized by the
existence of an energy gap for the creation of particle or hole excitations,
i.e., for the addition of particles to, or removal of particles from, the
system. In a periodic potential the bosons can move from one lattice site to
the next by quantum tunneling. The heights of the barriers can readily be
adjusted since they are proportional to the intensity of the laser beam. The
competition between the tunneling amplitude (kinetic energy) and the
atom-atom interactions controls the point of the phase transition. When the
tunneling coupling becomes large compared to the atom-atom interactions a
delocalized wavefunction minimizes the dominant kinetic energy and the
system undergoes a phase transition into the superfluid state. In such a
state the atom number per site is not fixed and the wavefunction exhibits
long-range phase coherence. Recent theoretical studies include a
delocalizing transition of BEC in optical lattices \cite{Kalo02}, the
superfluid density and the quasi-momentum distribution \cite{Roth03,Rey03},
the superfluid to Mott insulator transition with high filling factors
(multiband situation) \cite{Oost03}, commensurate-incommensurate transition 
\cite{Buch03}, as well as Monte Carlo simulations for the Bose-Hubbard
systems \cite{Kash02,Batr02}.

An advantage of an optical trap is that it liberates the spin degrees of
freedom and makes possible condensation of spinor bosons. For example, $%
^{23} $Na, $^{39}$K and $^{87}$Rb atoms (with nuclear spin $I=3/2$ and
electrons at $s$ orbits) at low temperatures behave as simple bosons with a
hyperfine spin $f=1$. Optically trapped BEC creates a possibility to study a
variety of phenomena in spinor many boson systems which posses extremely
rich physics. The insulator-superfluid transition in spinor BEC is one of
the possible novel phenomena. Recently the insulator-superfluid transition
of spin-1 bosons interacting antiferromagnetically in an optical lattice was
investigated theoretically \cite{Deml02,Tsuc02}, as well as the transition
in a multicomponent BEC system \cite{Chen03} and spin-2 Bose atoms \cite
{Hou03}. In this paper we study the phase transition of spin$-1$ bosons in
the presence of an external magnetic field. Magnetic field lifts a
degeneracy of the antiferromagnetic ground state and enriches the phase
diagram. We show that magnetic field results in a qualitatively new effect.
Depending on the field value the transition for bosons with
antiferromagnetic interparticle interaction may occur into different
superfluid phases with spins aligned along or opposite to the field
direction.

\section{Basic formalism}

We consider a dilute gas of trapped bosonic atoms with hyperfine spin $f=1$.
The system is described by the following Hamiltonian \cite{Law98}: $\hat H=%
\hat H_S+\hat H_A$, where the spin symmetric and asymmetric parts are 
\begin{equation}
\label{q1}\hat H_S=\sum_\alpha \int d^3r\hat \Psi _\alpha ^{+}\left( -\frac{%
\nabla ^2}{2M}+V_{{\rm tr}}\right) \hat \Psi _\alpha +\frac{\Lambda _s}2%
\sum_{\alpha ,\beta }\int \hat \Psi _\alpha ^{+}\hat \Psi _\beta ^{+}\hat 
\Psi _\alpha \hat \Psi _\beta d^3r, 
\end{equation}

$$
\hat H_A=\frac{\Lambda _a}2\int d^3r\left( \hat \Psi _1^{+}\hat \Psi _1^{+}%
\hat \Psi _1\hat \Psi _1+\hat \Psi _{-1}^{+}\hat \Psi _{-1}^{+}\hat \Psi
_{-1}\hat \Psi _{-1}+2\hat \Psi _1^{+}\hat \Psi _0^{+}\hat \Psi _1\hat \Psi
_0+2\hat \Psi _{-1}^{+}\hat \Psi _0^{+}\hat \Psi _{-1}\hat \Psi _0-2\hat \Psi
_1^{+}\hat \Psi _{-1}^{+}\hat \Psi _1\hat \Psi _{-1}+\right. 
$$
\begin{equation}
\label{q2}\left. +2\hat \Psi _0^{+}\hat \Psi _0^{+}\hat \Psi _1\hat \Psi
_{-1}+2\hat \Psi _1^{+}\hat \Psi _{-1}^{+}\hat \Psi _0\hat \Psi _0\right) , 
\end{equation}
here $\hat \Psi _\alpha $ ($\alpha =-1,0,1$) is the atomic field
annihilation operator associated with atoms in the hyperfine spin state $%
|f=1,m_f=\alpha >$, $V_{{\rm tr}}=V_0(\sin ^2kx+\sin ^2ky+\sin ^2kz)$ is the
optical lattice potential which is assumed to be the same for all three spin
components, $k=2\pi /\lambda $, $\lambda $ is the wavelength of the laser
light, $V_0$ is the tunable depth of the potential well. The coefficients $%
\Lambda _s$, $\Lambda _a$ are related to scattering lengths $a_0$ and $a_2$
of two colliding bosons with total angular momenta $0$ and $2$ by $\Lambda
_s=4\pi \hbar ^2(a_0+2a_2)/3M$ and $\Lambda _a=4\pi \hbar ^2(a_2-a_0)/3M$,
where $M$ is the mass of the atom \cite{Ho98}. For scattering of $^{23}$Na $%
a_2=54.7a_B$ and $a_0=49.4a_B$, where $a_B$ is the Bohr radius \cite{Burk98}%
. This suggests $\Lambda _a>0$ (antiferromagnetic interaction). For $^{87}$%
Rb $a_2=(107\pm 4)a_B$ and $a_0=(110\pm 4)a_B$ \cite{Ho98}, that is $\Lambda
_a<0$ and the interaction is ferromagnetic.

For single atoms the energy eigenstates are Bloch wave functions. An
appropriate superposition of Bloch states yields a set of Wannier functions
which are localized on the individual lattice sites. Expanding the field
operators in the Wannier basis

\begin{equation}
\label{q21}\hat \Psi _\alpha ({\bf r})=\sum_i\hat a_{\alpha i}w({\bf r-r}_i)%
\text{,} 
\end{equation}
where $\hat a_{\alpha i}$ correspond to the bosonic annihilation operators
on the $i$th lattice site, and keeping terms in first order in the hopping
matrix element, the Hamiltonian (\ref{q1}), (\ref{q2}) reduces to the
Bose-Hubbard Hamiltonian 
\begin{equation}
\label{q22}\hat H_S=-J\sum_{<i,j>}\sum_\alpha \hat a_{\alpha i}^{+}\hat a%
_{\alpha j}+\lambda _s\sum_i\hat N_i(\hat N_i-1), 
\end{equation}
\begin{equation}
\label{q8}\hat H_A=\lambda _a\sum_i(\hat a_1^{+}\hat a_1^{+}\hat a_1\hat a_1+%
\hat a_{-1}^{+}\hat a_{-1}^{+}\hat a_{-1}\hat a_{-1}-2\hat a_1^{+}\hat a%
_{-1}^{+}\hat a_1\hat a_{-1}+2\hat a_1^{+}\hat a_0^{+}\hat a_1\hat a_0+2\hat 
a_{-1}^{+}\hat a_0^{+}\hat a_{-1}\hat a_0+2\hat a_0^{+}\hat a_0^{+}\hat a_1%
\hat a_{-1}+2\hat a_1^{+}\hat a_{-1}^{+}\hat a_0\hat a_0), 
\end{equation}
where $\hat N_i=\sum_\alpha \hat a_{\alpha i}^{+}\hat a_{\alpha i}$ is the
number of atoms at lattice site $i$, $<i,j>$ stands for the nearest-neighbor
sites, $2\lambda _{s,a}=\Lambda _{s,a}\int d^3r|w|^4$ corresponds to the
strength of the on-site interaction between atoms and 
$$
J=-\int d^3rw^{*}({\bf r-r}_i)\left( -\frac{\nabla ^2}{2M}+V_{{\rm tr}%
}\right) w({\bf r-r}_j)>0 
$$
is the hopping matrix element between adjacent sites $i$,$j$. This describes
tunneling of atoms and is related to the kinetic energy. The parameter $J$
exponentially depends on the strength of periodic potential $V_0$ and can be
varied experimentally by several orders of magnitude. One should mention
that when number of particles in each well $N$ is large enough the
parameters $\lambda _{s,a}$ can be reduced by a factor $N^{-3/5}$. This
occurs when condensates in each well reach the Thomas-Fermi regime. However,
in experiments on 3D optical lattices $N\lesssim 100$ and the Thomas-Fermi
regime is usually not achieved. Higher filling factor can be realized in 2D
and 1D lattices \cite{Grei01,Orze01}. Also, equation (\ref{q21}) suggests
that atoms in different spin states are approximately described by the same
coordinate wave function which is the case when the spin symmetric
interaction is strong compared with the asymmetric part: $|\Lambda _s|\gg
|\Lambda _a|$ \cite{Law98}. This is relevant to experimental conditions for $%
^{23}$Na and $^{87}$Rb atoms \cite{Burk98,Ho98}. Introducing the operators

$$
\hat L_{+}=\sqrt{2}(\hat a_1^{+}\hat a_0+\hat a_0^{+}\hat a_{-1}),\quad \hat 
L_{-}=\sqrt{2}(\hat a_0^{+}\hat a_1+\hat a_{-1}^{+}\hat a_0),\quad \hat L_z=%
\hat a_1^{+}\hat a_1-\hat a_{-1}^{+}\hat a_{-1},\quad {\bf \hat L}^2=\hat L%
_{+}\hat L_{-}+\hat L_z^2-\hat L_z 
$$
which obey the angular momentum commutation relations $[\hat L_{+},\hat L%
_{-}]=2\hat L_z$, $[\hat L_z,\hat L_{\pm }]=\pm \hat L_{\pm }$ one can
obtain \cite{Law98} 
\begin{equation}
\label{q9}\hat H_A=\lambda _a\sum_i({\bf \hat L}_i^2-2\hat N_i). 
\end{equation}
The operators ${\bf \hat L}_i$ and $\hat L_{zi}$ commute with $\hat N_i$ and
each other.

To study the phase transitions, it is more convenient to perform
calculations in the grand-canonical ensemble, so we add the term with the
chemical potential $\mu $. We also include an external uniform magnetic
field $B$ along the $z$ axis \cite{Koas00}. Then the Hamiltonian becomes

\begin{equation}
\label{q11}\hat H=-J\sum_{<i,j>}\sum_\alpha \hat a_{\alpha i}^{+}\hat a%
_{\alpha j}+\sum_i\left[ \lambda _s\hat N_i(\hat N_i-1)+\lambda _a({\bf \hat 
L}_i^2-2\hat N_i)-\mu \hat N_i-b\hat L_{zi}\right] , 
\end{equation}
where $b=g\mu _BB$ which we assume to be positive, $g$ is a Land\'e factor
of an atom and $\mu _B$ is the Bohr magneton. As we will see later the most
interesting effects occur when the magnetic energy $g\mu _BB$ is less than
or comparable with the energy of spin asymmetric interaction. For such
values of magnetic field the energy of spin symmetric interaction is much
larger than $g\mu _BB$ and, hence, the description of atoms in different
spin states by the same coordinate wave function is justified for such
magnetic field.

The consistent mean-field theory we will use corresponds to the following
decomposition of the hopping terms \cite{Sach99,Oost01,Ferr02}: 
\begin{equation}
\label{q111}\hat a_{\alpha i}^{+}\hat a_{\alpha j}=<\hat a_{\alpha i}^{+}>%
\hat a_{\alpha j}+\hat a_{\alpha i}^{+}<\hat a_{\alpha j}>-<\hat a_{\alpha
i}^{+}><\hat a_{\alpha j}>=\psi _\alpha (\hat a_{\alpha i}^{+}+\hat a%
_{\alpha j})-\psi _\alpha ^2, 
\end{equation}
where $\psi _\alpha =<\hat a_{\alpha i}^{+}>=<\hat a_{\alpha i}>$ is the
three component superfluid order parameter. In such a decomposition we omit
the higher-order fluctuations $(\hat a_{\alpha i}^{+}-\psi _\alpha )(\hat a%
_{\alpha j}-\psi _\alpha )$. Also we assume the order parameter to be real
and neglect the Josephson-type tunneling term. Then Eq. (\ref{q11}) yields 
\begin{equation}
\label{q112}\hat H=zJN_s\sum_\alpha \psi _\alpha ^2+\sum_i\left[ \lambda _s%
\hat N_i(\hat N_i-1)+\lambda _a({\bf \hat L}_i^2-2\hat N_i)-\mu \hat N_i-b%
\hat L_{zi}-zJ\sum_\alpha \psi _\alpha (\hat a_{\alpha i}^{+}+\hat a_{\alpha
i})\right] , 
\end{equation}
where $z=2d$ is the number of nearest-neighbor sites, $d$ is the space
dimension and $N_s$ is the total number of lattice sites. This Hamiltonian
is diagonal with respect to the site index $i$, so one can use an effective
onsite Hamiltonian $\hat H_i^{{\rm eff}}$ 
\begin{equation}
\label{q113}\hat H_i^{{\rm eff}}=\lambda _s^{}\hat N_i(\hat N_i-1)+\lambda
_a^{}({\bf \hat L}_i^2-2\hat N_i)-\mu \hat N_i-b\hat L_{zi}+zJ\sum_\alpha
\psi _\alpha ^2-zJ\sum_\alpha \psi _\alpha (\hat a_{\alpha i}^{+}+\hat a%
_{\alpha i}). 
\end{equation}
One should note that the simple mean-field approximation we use may not
catch some interesting features discussed in \cite{Deml02} for a field-free
transition. However, our simple approximation is good enough to describe a
new effect we predict for a transition in magnetic field and, hence, there
is no need to consider more complicated theories.

\section{Perturbation expansion over the superfluid order parameter}

We will treat the last two terms in Eq. (\ref{q113}) as a perturbation. For
the second order phase transition the order parameter $\psi _\alpha $
continuously changes from zero (in insulator phase) to a finite value (in
superfluid phase). Hence, in the vicinity of phase transition the order
parameter is infinitesimally small and, therefore, to find the transition
point the last two terms in Eq. (\ref{q113}) can be treated as a
perturbation. In zero order approximation (we drop the site index $i$ since
the effective Hamiltonian is diagonal with respect to the lattice sites)

\begin{equation}
\label{q12}\hat H_0^{{\rm eff}}=\lambda _s^{}\hat N(\hat N-1)+\lambda _a^{}(%
{\bf \hat L}^2-2\hat N)-\mu \hat N-b\hat L_z. 
\end{equation}
The eigenstates of $\hat H_0^{{\rm eff}}$ are states with defined total
number of atoms in the site $N$ (an integer number), the total spin per site 
$l$ and its projection $m$ along the magnetic field. We denote the
eigenstates as $|N,l,m>$. The corresponding eigenvalues are 
\begin{equation}
\label{q122}E_0=\lambda _sN(N-1)+\lambda _a[l(l+1)-2N]-\mu N-bm. 
\end{equation}
For $b>0$ the state with $m=l$ corresponds to the lowest energy. We assume
that all the atoms in the well are in the same orbital state, which is the
ground state for the confining potential. Then the symmetry of the bosonic
wave function and the structure of the operator ${\bf \hat L}$ enforces a
constraint on $l.$ The allowable values of $l$ are $l=0,2,4,...N$ if $N$ is
even and $l=1,3,5,...N$ if $N$ is odd (also $m=0,\pm 1,\pm 2,\ldots ,\pm l$
for any allowable $l$). This result is known in cavity QED; the details of
the proof are provided in Appendix A of Ref. \cite{Wu96}. The eigenfunctions 
$|N,l,m>$ in terms of the Fock states $|n_1,n_0,n_{-1}>$ with defined number
of particles $n_\alpha $ with the spin projection $\alpha $ are given by 
\begin{equation}
\label{q124}|N,l,m>=\sum_kA_k|k,N-2k-m,k+m>, 
\end{equation}
where $A_k$ satisfy the recursion relation%
$$
A_{k+1}\sqrt{(k+1)(k+m+1)(N-2k-m-1)(N-2k-m)}+A_{k-1}\sqrt{%
k(k+m)(N-2k-m+1)(N-2k-m+2)}+\text{ } 
$$

$$
+A_k[k(N-2k-m+1)+(N-2k-m)(k+m+1)]=[l(l+1)-m(m+1)]A_k/2\text{. } 
$$
In some particular cases this equation has simple solutions. E.g., if $m=l$
the coefficients are%
$$
A_k=(-1)^k\sqrt{\frac{(k+l)!}{k!}}\sqrt{\frac{(N-l-2k-1)!!}{(N-l-2k)!!}}A_0%
\text{,} 
$$
while for $m=l-1$ 
$$
A_k=(-1)^k\sqrt{\frac{(k+l)!}{k!}}\sqrt{\frac{(N-l-2k)!!}{(N-l-2k-1)!!}}A_0. 
$$
The coefficient $A_0$ can be obtained from the normalization condition $%
\sum_kA_k^2=1$.

\subsection{Mott ground state of spin-1 bosons}

The ground state of $\hat H_0^{{\rm eff}}$ depends on the relation between $%
\lambda _s$, $\lambda _a$, $b$ and $\mu $. It is determined by minimizing
the energy (\ref{q122}) with the constraint $N+l={\rm even}$. We assume $%
\lambda _s>0$. When $\lambda _a^{}<0$ the ground state is $|N,N,N>$, it is a
ferromagnetic state in which all bosons occupy the $m=1$ state. The number
of particles $N$ is determined from the condition 
\begin{equation}
\label{np2}2(N-1)(\lambda _s^{}+\lambda _a^{})<\mu +b<2N(\lambda
_s^{}+\lambda _a^{}). 
\end{equation}

When $\lambda _a^{}>0$ the nature of the ground state depends on magnetic
field. For $b=0$ the ground state is antiferromagnetic. However, the number
of particles per site depends on the relation between $\lambda _s$ and $%
\lambda _a$. There are two possibilities:

1) $\lambda _s>2\lambda _a$, then the particle number is determined from 
\begin{equation}
\label{q123}\left\{ 
\begin{array}{c}
2(N-1)\lambda _s-4\lambda _a^{}<\mu <2N\lambda _s,\quad N=0,2,4,...,\quad
l=0 \\ 
2(N-1)\lambda _s<\mu <2N\lambda _s-4\lambda _a^{},\quad N=1,3,5,...,\quad
l=1 
\end{array}
\right. . 
\end{equation}
The number of particles at each site increases by +1 any time the chemical
potential $\mu $ crosses the points $0$, $2\lambda _s-4\lambda _a^{}$, $%
4\lambda _s$, $6\lambda _s-4\lambda _a^{}$, $...$

2) $\lambda _s<2\lambda _a$, then states with even $N$ are the only
possibility and 
\begin{equation}
\label{q1231}(2N-3)\lambda _s-2\lambda _a^{}<\mu <(2N+1)\lambda _s-2\lambda
_a^{},\quad N=0,2,4,...,\quad l=0. 
\end{equation}
The particle number increases by +2 each time the chemical potential $\mu $
crosses the points $\lambda _s-2\lambda _a^{}$, $5\lambda _s-2\lambda _a^{}$%
, $9\lambda _s-2\lambda _a^{}$, $13\lambda _s-2\lambda _a^{}$, $...$. For $%
\lambda _a^{}>0$ and $b=0$ the insulator-superfluid transition has been
studied in detail in \cite{Tsuc02}.

For $\lambda _a^{},b>0$ and given $N$ the number $l$ is determined by the
parity of $N$ and the condition%
$$
(2l-1)\lambda _a<b<(2l+3)\lambda _a. 
$$
When $b>\lambda _a(2N-1)$ the ground state is ferromagnetic with $l=N$. Let
us define $l_{\max }$ so that $\lambda _a(2l_{\max }-1)<b<\lambda
_a(2l_{\max }+1)$. Then for $b>0$ the particle number is determined from (we
assume $\lambda _s>\lambda _a>0$)%
$$
N\leq l_{\max }-1:\quad 2(N-1)(\lambda _s+\lambda _a)<\mu +b<2N(\lambda
_s+\lambda _a), 
$$
$$
N\geq l_{\max }:\quad 2(N-1)\lambda _s-2\lambda _a+(-1)^{N+l_{\max
}}(2l_{\max }\lambda _a^{}-b)<\mu <2N\lambda _s-2\lambda _a-(-1)^{N+l_{\max
}}(2l_{\max }\lambda _a^{}-b). 
$$

\subsection{Phase diagram}

We calculate the phase boundary between the insulator and the superfluid
states for $b>0$ using second-order perturbation theory. The idea is to find
the energy as a function of the superfluid order parameter $\psi _\alpha $: $%
E=E_0+B_1\psi _1^2+B_0\psi _0^2+B_{-1}\psi _{-1}^2+\ldots $, where
coefficients $B_\alpha $ depend on the system's parameters. In the insulator
phase all $B_\alpha >0$ and, hence, $\psi _\alpha =0$ minimizes the energy.
Phase transition into a superfluid state occurs when one of the coefficients 
$B_\alpha $ becomes negative. In Appendix A we calculate the necessary
matrix elements of the perturbation operator between the ground state $%
|N,l,l>$ and excited states of $\hat H_0^{{\rm eff}}$. Here we list the
nonzero matrix elements:%
$$
<N+1,l+1,l|\hat a_0^{+}|N,l,l>=\sqrt{\frac{N+l+3}{2l+3}},\quad <N-1,l+1,l|%
\hat a_0|N,l,l>=\sqrt{\frac{N-l}{2l+3}}, 
$$
$$
<N+1,l+1,l+1|\hat a_1^{+}|N,l,l>=\sqrt{\frac{(l+1)(N+l+3)}{2l+3}},\quad
<N-1,l+1,l-1|\hat a_1|N,l,l>=-\sqrt{\frac{N-l}{(2l+3)(2l+1)}},\quad 
$$
$$
<N-1,l-1,l-1|\hat a_1|N,l,l>=\sqrt{\frac{l(N+l+1)}{2l+1}},\quad <N+1,l+1,l-1|%
\hat a_{-1}^{+}|N,l,l>=\sqrt{\frac{N+l+3}{(2l+3)(2l+1)}}, 
$$
$$
<N+1,l-1,l-1|\hat a_{-1}^{+}|N,l,l>=-\sqrt{\frac{l(N-l+2)}{2l+1}},\quad
<N-1,l+1,l+1|\hat a_{-1}|N,l,l>=-\sqrt{\frac{(N-l)(l+1)}{2l+3}}. 
$$
The second order perturbation theory results in the following expression for
the energy correction to the $|N,l,l>$ state in terms of the superfluid
order parameter $\psi _\alpha $%
$$
E=\lambda _sN(N-1)+\lambda _a[l(l+1)-2N]-\mu N-bl+zJ\sum_\alpha \psi _\alpha
^2- 
$$
$$
-\frac{(zJ)^2\psi _1^2}{(2l+3)(2l+1)}\left[ \frac{(l+1)(N+l+3)(2l+1)}{%
2N\lambda _s+2l\lambda _a-b-\mu }+\frac{N-l}{2(1-N)\lambda _s+2(l+2)\lambda
_a+b+\mu }+\frac{l(N+l+1)(2l+3)}{2(1-N)\lambda _s-2(l-1)\lambda _a+b+\mu }%
\right] - 
$$
$$
-\frac{(zJ)^2\psi _0^2}{2l+3}\left[ \frac{N+l+3}{2N\lambda _s+2l\lambda
_a-\mu }+\frac{N-l}{2(1-N)\lambda _s+2(l+2)\lambda _a+\mu }\right] - 
$$
\begin{equation}
\label{c17}-\frac{(zJ)^2\psi _{-1}^2}{(2l+3)(2l+1)}\left[ \frac{N+l+3}{%
2N\lambda _s+2l\lambda _a+b-\mu }+\frac{l(N-l+2)(2l+3)}{2N\lambda
_s-2(l+1)\lambda _a+b-\mu }+\frac{(N-l)(l+1)(2l+1)}{2(1-N)\lambda
_s+2(l+2)\lambda _a-b+\mu }\right] . 
\end{equation}
The leading correction to the energy is quadratic in $\psi _\alpha $ which
demonstrates close analogy of the approach with Schrieffer-Wolff
transformation. The idea of such canonical transformation is to eliminate
the perturbation term in the Hamiltonian to first order \cite{Schr66}. Phase
transition into a superfluid state with $\psi _\alpha \neq 0$ occurs when
the coefficient in front of $\psi _\alpha ^2$ becomes negative. For $b=0$
and $\lambda _a^{}>0$ $\,$the superfluid transition occurs into a polar (spin%
$-0$) state; the result is valid for both even and odd number of particles
per site \cite{Tsuc02}. In such a transition the order parameter $\psi _0$
becomes nonzero.

When $b>0$ the situation changes substantially. One can see from Eq. (\ref
{c17}) that as soon as the ground state corresponds to $l\neq 0$ the
coefficient in front of $\psi _1^2$ in the expression for $E$ becomes
smaller than the coefficient in front $\psi _0^2$. At the same time the
terms containing $\psi _1$ and $\psi _{-1}$ compete with each other.
Depending on the value of $b$ and the chemical potential (particle number)
the system undergoes a transition into a superfluid state with $\psi _1\neq
0 $ or $\psi _{-1}\neq 0$.

The coefficient in front of $\psi _1^2$ becomes zero when%
$$
\frac{(2l+3)(2l+1)}{zJ}=\frac{(l+1)(N+l+3)(2l+1)}{2N\lambda _s+2l\lambda
_a-b-\mu }+\frac{N-l}{2(1-N)\lambda _s+2(l+2)\lambda _a+b+\mu }+ 
$$
\begin{equation}
\label{c18}+\frac{l(N+l+1)(2l+3)}{2(1-N)\lambda _s-2(l-1)\lambda _a+b+\mu }, 
\end{equation}
while for the coefficient in front of $\psi _{-1}^2$ the condition is%
$$
\frac{(2l+3)(2l+1)}{zJ}=\frac{(N+l+3)}{2N\lambda _s+2l\lambda _a+b-\mu }+%
\frac{(N-l)(l+1)(2l+1)}{2(1-N)\lambda _s+2(l+2)\lambda _a-b+\mu }+ 
$$
\begin{equation}
\label{c19}+\frac{l(N-l+2)(2l+3)}{2N\lambda _s-2(l+1)\lambda _a+b-\mu }. 
\end{equation}
The lowest value of $J$ from Eqs. (\ref{c18}), (\ref{c19}) determines the
point of phase transition. In Fig. 1 we plot the phase boundary in the $%
J-\mu $ plane for the insulator-superfluid transition in a magnetic field.
In our estimates we take $\lambda _a=0.4\lambda _s$ and $b=4.25\lambda _a$.
For $J=0$ (no tunneling) the system is in the ground state for a single
well. The number of particles per site $N$ as well as the total spin $l$
depend on the chemical potential $\mu $. Solid line corresponds to a
transition into superfluid phase with $\psi _1\neq 0$, while along the dash
line the transition occurs into a state with $\psi _{-1}\neq 0$. The phase
boundaries between different superfluid phases with either non-zero $\psi _1$
or $\psi _{-1}$ are beyond the scope of the present paper and are not shown.
In Fig. 2 we plot the phase diagram for $\lambda _a=0.4\lambda _s$ and $%
b=6\lambda _a$. The possibility of the phase transition into different
superfluid states can be understood as follows. In the insulator phase
particle permutation symmetry imposes the constraint $N+l={\rm even}$; the
parity of $l$ is fixed by the parity of $N$. However, the interaction energy 
$\lambda _al(l+1)-bl$ can be lower for $l$ with the opposite parity. The
symmetry constraint is relaxed in the superfluid phase. Due to the
appearance of a coherent superfluid component the particle number per site
is no longer fixed. Effectively the superfluidity removes the restriction on
the $l$'s parity and, as a result, the superfluid order parameter may
correspond to different projections of spins along the magnetic field in
order to decrease the excessive interaction energy. The effect takes place
for small magnetic field or large particle number per site when $\lambda
_a(2N-1)>b$. The coefficient $\lambda _a$ can be estimated as $\lambda
_a=(\pi ^2/3)E_R[(a_2-a_0)/\lambda ](V_0/E_R)^{3/4}$, where $E_R=2\pi
^2\hbar ^2/M\lambda ^2$ is the recoil energy \cite{Deml02}. For $^{23}$Na
atoms $a_2-a_0\approx 5.3a_B$ and typical experimental parameters \cite
{Stam98} $\lambda =0.985{\rm \mu m}$, $E_R=4.4\times 10^{-7}$K, $V_0\sim
12E_R$, $N\approx 3$ we obtain the following limitation on the magnetic
field $B<10^{-4}{\rm G}$. Since the current capability of magnetic shielding
can reach $10^{-5}{\rm G}$, the low magnetic field regime is attainable.
Experimentally the transition into different superfluid phases may be
detected, for example, by creating a weak gradient of magnetic field. In
such a case the superfluids with different magnetization will move in
opposite directions.

For $b>\lambda _a(2N-1)$ or $\lambda _a^{}<0$ the ground state is
ferromagnetic with $l=N$. In this case the coefficient in front of $\psi
_1^2 $ in the expression for the $E$ is smallest among the coefficients in
front of $\psi _1^2,$ $\psi _0^2,$ $\psi _{-1}^2$. This means that the order
parameter $\psi _1$ always appears first and the system undergoes a
transition into ferromagnetic superfluid with $\psi _1\neq 0$. The
transition occurs when 
\begin{equation}
\label{c20}zJ=\frac{(2N(\lambda _s+\lambda _a)-b-\mu )(b+\mu -2(N-1)(\lambda
_s+\lambda _a))}{2(\lambda _s+\lambda _a)+b+\mu },\quad 2(N-1)(\lambda
_s+\lambda _a)<b+\mu <2N(\lambda _s+\lambda _a). 
\end{equation}
In Figs. 1 and 2 the ferromagnetic state is realized for small particle
numbers per site ($N<2$ and $N<3$).

One should mention that we neglected possible presence of slowly varying (at
the length scale of the lattice period) component of the trapping potential.
Such potential effectively provides a scan over $\mu $ of the phase diagram
at fixed values of $J/\lambda _s$ and $\lambda _a/\lambda _s$. Also at
higher value of the tunneling amplitude $J$ additional phase transitions
between different superfluid states may occur. Their description requires
estimate of the next order terms in the energy and lies beyond the scope of
our paper. One of the interesting question for future studies is
superfluidity of ``holes'' which might occur when some sites remain empty
after filling the optical lattice by atoms. Also, thermal fluctuations may
become important at nonzero temperature and destroy long-range phase
coherence of superfluid. As a result, the transition can first occur into a
conductor phase (with no long-range coherence) and only later into a
superfluid state.

This work was supported by NASA, Grant No. NAG8-1427.

\appendix

\section{Calculation of matrix elements}

In the presence of magnetic field the ground state of zero-order Hamiltonian
(\ref{q12}) is $|N,l,l>$. Let us consider matrix elements of the
perturbation operator $\hat a_\alpha ^{+}+\hat a_\alpha $ between the ground
state and the other eigenstates. To calculate matrix elements we consider
the eigenstates of the Hamiltonian (\ref{q12}) in a form \cite{Ho00} 
\begin{equation}
\label{c1}|N,l,l>=\frac 1{\sqrt{f(N,l)}}(\hat a_1^{+})^l(\hat \Theta
^{+})^{(N-l)/2}|vac>, 
\end{equation}
where $\hat \Theta ^{+}=\hat a_0^{+2}-2\hat a_1^{+}\hat a_{-1}^{+}$ and $f$
is a normalization factor%
$$
f(N,l)=l!\left( \frac{N-l}2\right) !2^{(N-l)/2}\frac{(N+l+1)!!}{(2l+1)!!}. 
$$
States with lower magnetic quantum numbers can be obtained by operating $%
\hat L_{-}=\sqrt{2}(\hat a_0^{+}\hat a_1+\hat a_{-1}^{+}\hat a_0)$, $\hat L%
_{+}=\sqrt{2}(\hat a_1^{+}\hat a_0+\hat a_0^{+}\hat a_{-1})$ to $|N,l,l>$
and using $\hat L_{-}|N,l,m>=\sqrt{(l+m)(l-m+1)}|N,l,m-1>$, $\hat L%
_{+}|N,l,m>=\sqrt{(l+m+1)(l-m)}|N,l,m+1>$. Another useful operator is $\hat L%
_z=\hat a_1^{+}\hat a_1-\hat a_{-1}^{+}\hat a_{-1}$, $\hat L%
_z|N,l,m>=m|N,l,m>$. We note the properties 
$$
[\hat L_{\pm },\hat \Theta ^{+}]=0,\quad [\hat L_z,\hat \Theta ^{+}]=0,\quad
[\hat L_{-},(\hat a_1^{+})^l]=\sqrt{2}l\hat a_0^{+}(\hat a%
_1^{+})^{l-1},\quad [\hat L_{-},\hat a_0^{+}]=\sqrt{2}\hat a_{-1}^{+}, 
$$
$$
\hat \Theta ^{+}|N,l,l>=\sqrt{(N-l+2)(N+l+3)}|N+2,l,l>,\quad \hat L_{\pm
}|vac>=0,\quad \hat L_z|vac>=0, 
$$
$$
[\hat a_1,(\hat \Theta ^{+})^n]=-2n\hat a_{-1}^{+}(\hat \Theta
^{+})^{n-1},\quad [\hat a_0,(\hat \Theta ^{+})^n]=2n\hat a_0^{+}(\hat \Theta
^{+})^{n-1},\quad [\hat a_{-1},(\hat \Theta ^{+})^n]=-2n\hat a_1^{+}(\hat 
\Theta ^{+})^{n-1}, 
$$
$$
[\hat L_{+},\hat a_{-1}^{+}]=\sqrt{2}\hat a_0^{+},\quad [\hat L_{+},\hat a%
_1]=-\sqrt{2}\hat a_0,\quad [\hat L_z,\hat a_{-1}^{+}]=-\hat a%
_{-1}^{+},\quad [\hat L_z,\hat a_1]=-\hat a_{-1}, 
$$
where $[\hat a,\hat b]$ stands for the commutator of $\hat a$ and $\hat b$.
Then using (\ref{c1}) we obtain 
$$
\sqrt{f(N-1,l-1)}\hat a_0^{+}|N-1,l-1,l-1>=\hat a_0^{+}(\hat a_1^{+})^{l-1}(%
\hat \Theta ^{+})^{(N-l)/2}|vac>=\frac 1{\sqrt{2}l}[\hat L_{-},(\hat a%
_1^{+})^l](\hat \Theta ^{+})^{(N-l)/2}|vac>= 
$$
$$
=\frac 1{\sqrt{2}l}\hat L_{-}(\hat a_1^{+})^l(\hat \Theta
^{+})^{(N-l)/2}|vac>=\frac{\sqrt{f(N,l)}}{\sqrt{2}l}\hat L_{-}|N,l,l>=\frac{%
\sqrt{f(N,l)}}{\sqrt{l}}|N,l,l-1>\text{,} 
$$
or 
\begin{equation}
\label{c2}\hat a_0^{+}|N,l,l>=\sqrt{\frac{N+l+3}{2l+3}}|N+1,l+1,l>. 
\end{equation}
Further%
$$
\hat a_1^{+}|N,l,l>=\frac 1{\sqrt{f(N,l)}}(\hat a_1^{+})^{l+1}(\hat \Theta
^{+})^{(N-l)/2}|vac>=\sqrt{\frac{f(N+1,l+1)}{f(N,l)}}|N+1,l+1,l+1>= 
$$
\begin{equation}
\label{c3}=\sqrt{\frac{(l+1)(N+l+3)}{2l+3}}|N+1,l+1,l+1>. 
\end{equation}
$$
\hat a_{-1}|N,l,l>=\frac 1{\sqrt{f(N,l)}}(\hat a_1^{+})^l\hat a_{-1}(\hat 
\Theta ^{+})^{(N-l)/2}|vac>=-\frac{(N-l)}{\sqrt{f(N,l)}}(\hat a_1^{+})^{l+1}(%
\hat \Theta ^{+})^{(N-l-2)/2}|vac>= 
$$
\begin{equation}
\label{c5}=-(N-l)\sqrt{\frac{f(N-1,l+1)}{f(N,l)}}|N-1,l+1,l+1>=-\sqrt{\frac{%
(N-l)(l+1)}{2l+3}}|N-1,l+1,l+1>. 
\end{equation}
$$
\hat a_0|N,l,l>=\frac 1{\sqrt{f(N,l)}}(\hat a_1^{+})^l\hat a_0(\hat \Theta
^{+})^{(N-l)/2}|vac>=\frac{(N-l)}{\sqrt{f(N,l)}}(\hat a_1^{+})^l\hat a_0^{+}(%
\hat \Theta ^{+})^{(N-l-2)/2}|vac>= 
$$
\begin{equation}
\label{c6}=(N-l)\sqrt{\frac{f(N-2,l)}{f(N,l)}}\hat a_0^{+}|N-2,l,l>=\frac{%
\sqrt{N-l}}{\sqrt{2l+3}}|N-1,l+1,l>. 
\end{equation}
Now let us find $\hat a_{-1}^{+}|N,l,l>$. Using%
$$
\sqrt{2}\hat a_0^{+}|N,l,l>=\hat L_{+}\hat a_{-1}^{+}|N,l,l>, 
$$
we obtain 
\begin{equation}
\label{c7}\hat a_{-1}^{+}|N,l,l>=\sqrt{\frac{N+l+3}{(2l+3)(2l+1)}}%
|N+1,l+1,l-1>+B|N+1,l-1,l-1>, 
\end{equation}
where $B$ is a coefficient. Other terms such as $|N+1,m,m>$ do not enter Eq.
(\ref{c7}). This can be proven by applying $\hat L_z$ to Eq. (\ref{c7}) and
using the property $\hat L_z\hat a_{-1}^{+}|N,l,l>=(l-1)\hat a%
_{-1}^{+}|N,l,l>$. To find the coefficient $B$ one can apply $\hat a_{-1}$
to Eq. (\ref{c7}) and consider the matrix element $<N,l,l|\hat a_{-1}\hat a%
_{-1}^{+}|N,l,l>=1+<N,l,l|\hat a_{-1}^{+}\hat a%
_{-1}|N,l,l>=1+(N-l)(l+1)/(2l+3)$. As a result, we obtain $B=-\sqrt{%
l(N-l+2)/(2l+1)}$. Hence 
\begin{equation}
\label{c71}\hat a_{-1}^{+}|N,l,l>=\sqrt{\frac{N+l+3}{(2l+3)(2l+1)}}%
|N+1,l+1,l-1>-\sqrt{\frac{l(N-l+2)}{2l+1}}|N+1,l-1,l-1>. 
\end{equation}
Finally, let us calculate $\hat a_1|N,l,l>$. Using%
$$
-\sqrt{2}\hat a_0|N,l,l>=\hat L_{+}\hat a_1|N,l,l>, 
$$
we find%
$$
\hat a_1|N,l,l>=-\sqrt{\frac{N-l}{(2l+3)(2l+1)}}%
|N-1,l+1,l-1>+C|N-1,l-1,l-1>. 
$$
The coefficient $C$ can be obtained by applying $\hat a_1^{+}$ and
considering the matrix element $<N,l,l|\hat a_1^{+}\hat a_1|N,l,l>$.
Ultimately 
\begin{equation}
\label{c8}\hat a_1|N,l,l>=-\sqrt{\frac{N-l}{(2l+3)(2l+1)}}|N-1,l+1,l-1>+%
\sqrt{\frac{l(N+l+1)}{2l+1}}|N-1,l-1,l-1>. 
\end{equation}

\begin{figure}
\bigskip
\centerline{\epsfxsize=0.48\textwidth\epsfysize=0.48\textwidth
\epsfbox{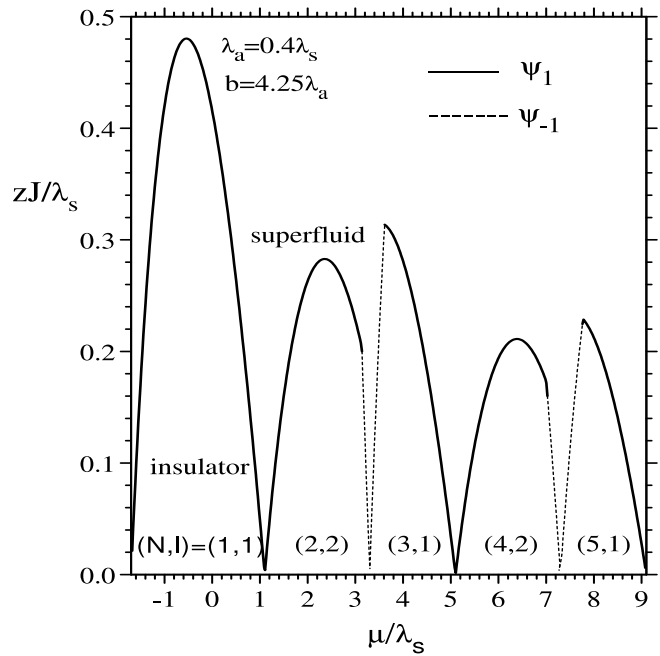}}

\vspace{0.5cm}

\caption{Phase diagram of spin-1 bosons in an optical lattice in $J-\mu $ plane
for a fixed value of magnetic field $b=4.25\lambda _a$ and
$\lambda _a=0.4\lambda _s$. Solid line corresponds to a transition from an
insulator into superfluid phase with $\psi _1\neq 0$, while along the dash line
the transition occurs into a state with $\psi _{-1}\neq 0$. }

\label{fig1}
\end{figure}

\begin{figure}
\bigskip
\centerline{\epsfxsize=0.6\textwidth\epsfysize=0.6\textwidth
\epsfbox{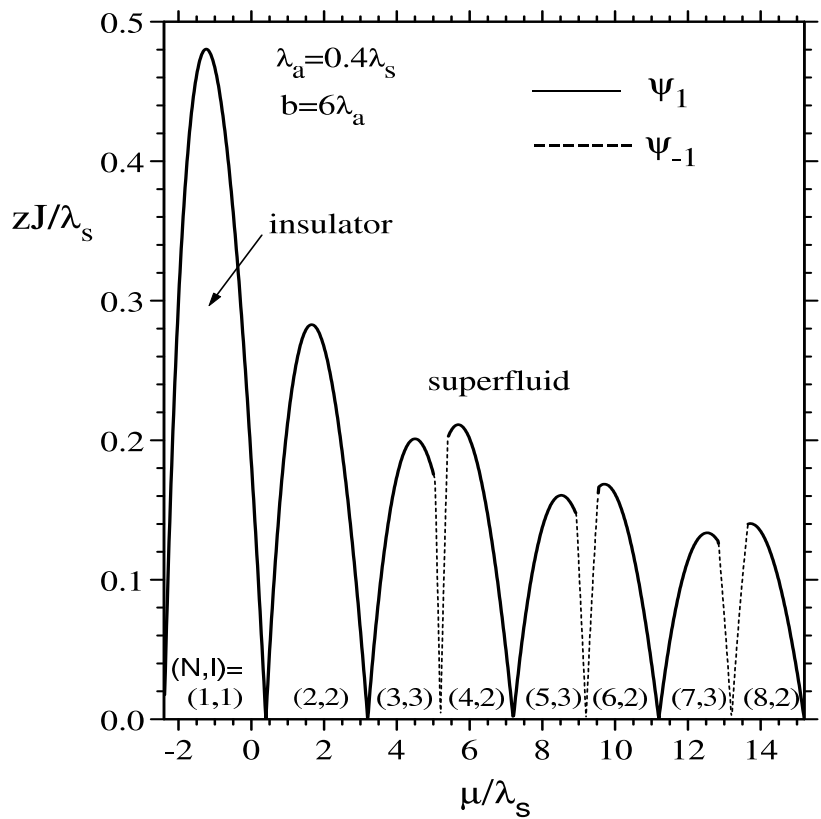}}

\vspace{0.5cm}

\caption{The same as in Fig. 1, but for $b=6\lambda _a$ and
$\lambda_a=0.4\lambda _s$.}

\label{fig2}
\end{figure}

\end{document}